\documentclass{emulateapj}
\usepackage[colorlinks,urlcolor=blue,citecolor=blue,linkcolor=blue]{hyperref} 
\usepackage{macros,graphicx,natbib}

\begin{document}

\title{All Transients, All the Time: \\ Real-Time Radio Transient Detection with Interferometric Closure Quantities}

\author{Casey J. Law}

\author{Geoffrey C. Bower}

\affil{Department of Astronomy and Radio Astronomy Lab, University of California, Berkeley, CA}

\keywords{Instrumentation: interferometers, Techniques: interferometric, pulsars: individual (Crab, B0329+54)}

\begin{abstract}
We demonstrate a new technique for detecting radio transients based on interferometric closure quantities. The technique uses the bispectrum, the product of visibilities around a closed-loop of baselines of an interferometer. The bispectrum is calibration independent, resistant to interference, and computationally efficient, so it can be built into correlators for real-time transient detection. Our technique could find celestial transients anywhere in the field of view and localize them to arcsecond precision. At the Karl G. Jansky Very Large Array (VLA), such a system would have a high survey speed and a 5$\sigma$\ sensitivity of 38 mJy on 10 ms timescales with 1 GHz of bandwidth. The ability to localize dispersed millisecond pulses to arcsecond precision in large volumes of interferometer data has several unique science applications. Localizing individual pulses from Galactic pulsars will help find X-ray counterparts that define their physical properties, while finding host galaxies of extragalactic transients will measure the electron density of the intergalactic medium with a single dispersed pulse. Exoplanets and active stars have distinct millisecond variability that can be used to identify them and probe their magnetospheres. We use millisecond time scale visibilities from the Allen Telescope Array (ATA) and VLA to show that the bispectrum can detect dispersed pulses and reject local interference. The computational and data efficiency of the bispectrum will help find transients on a range of time scales with next-generation radio interferometers.
\end{abstract}

\section{Introduction}
Since the discovery of the first pulsar, radio observations have steadily revealed a wide range of fast ($\lesssim1$s) transient phenomena. Pulsars \citep{1968Natur.217..709H,1982Natur.300..615B,2006Natur.442..892C}, planets and active stars \citep{2007ApJ...663L..25H,2007AA...475..359G}, and peculiar new transients \citep{2006Natur.439..817M,2007Sci...318..777L} all have distinct radio phenomenology related to the physics of the host and the medium through which the signal propagates.

While large, single-dish telescopes have pioneered the study of fast radio transients, the nature of their design has several limits. Single-dish telescopes can only localize an individual pulse to somewhere in its field of view \citep[typically arcminutes;][]{2001MNRAS.328...17M}. Timing of periodic emission can localize very precisely, but individual pulses can't be localized well enough to easily identify optical or X-ray counterparts. Since interferometers simultaneously have a large collecting area and a large field of view, they survey more efficiently than single-dish telescopes. Survey speed scales roughly the square of the product of the number of dishes and their diameter; this predicts that the VLA surveys almost 40 times faster than the Green Bank Telescope, assuming all other factors are equal. Despite this potential, next-generation radio interferometers (VLA, PAPER, LOFAR, ASKAP, and MeerKAT) have fast transients observing modes that are essentially identical to the single-dish concept: point one or more beams on the sky and search for pulses \citep{2011ApJ...734...20M,2011arXiv1109.2659S}.

In principle, interferometers have much more information available to them because they detect the electromagnetic wave over a distributed array of receivers. Instead of forming a beam, interferometers can measure visibilities by correlating signals across the array. Visibilities contain spatial information that can reject interference and localize a source to \emph{arcsecond} precision over a large field of view. Traditionally, visibilities haven't been used in this way because it was difficult to run correlators this fast and the data volume was too large to handle or efficiently search for pulses \citep{2011ApJ...742...12L}. As a result, a standard radio interferometer averages data to second time scales, losing nearly all information about millisecond variability in the sky. Even on these longer time scales, imaging algorithms are complex and subject to subtle statistical biases \citep{1998AJ....115.1693C}. The challenge of data rates and volumes is expected to grow with next-generation interferometers, which will have more antennas, wider bandwidths, and larger fields of view \citep{2011ApJ...739L...1P}.

Here, we present a technique that reduces the challenge of detecting transients in visibilities to the much simpler single-dish problem, while maintaining the utility of visibilities. The technique is based on the ``bispectrum'', a quantity formed from visibilities that is sensitive to a point source anywhere in the field of view \citep{1987AA...180..269C,1989AJ.....98.1112K,1995AJ....109.1391R}. Differencing visibilities makes it possible to interpret the bispectrum as a statistic for transients anywhere in the field of view on time scales from milliseconds to seconds. We describe an algorithm for real-time, arcsecond localization of transients using raw visibilities from a radio interferometer.

The theory of closure quantities and their application to transients is presented in \S \ref{theory}. In \S \ref{demo} we test the theoretical predictions with millisecond time scale visibilities from the VLA and the ``Pocket Correlator'' (PoCo) instrument at the ATA \citep{2011ApJ...742...12L}. As shown in \S \ref{demand}, this technique has a moderate computational demand and can reduce the flow of data to a manageable size. We conclude with thoughts about new science accessible with this technique and how it can address the growing challenge of big data in the study of transients with radio interferometers.

\section{Theory of Transient Detection with the Bispectrum}
\label{theory}
Closure quantities are the combination of visibilities from antennas that form a closed loop \citep{1987AA...180..269C}. An example of a closure quantity is the closure phase or triple phase, which is the sum of the visibility phase from three baselines, $\phi_{\rm{ijk}} = \phi_{ij} + \phi_{jk} + \phi_{ki}$. This combination is useful because it is independent of phase errors associated with any individual antenna. Closure quantities have largely been developed for very long baseline interferometry, because they make it possible to detect a source even if phase calibration is unstable \citep{1995AJ....109.1391R}. An interesting corollary is that the triple phase responds in the same way to a point source anywhere in the field of view.

A related closure quantity is the product of visibilities on a closed loop, called the closure triple or bispectrum. The bispectrum can be written as $b_{ijk} = a_{ij} \, a_{jk} \, a_{ki} \, e^{i\phi_{\rm{ijk}}}$, where $a_{ij}$ are the amplitudes of visibilities for baseline $ij$. In the absence of signal, the bispectra have a mean at the origin of the real-imaginary plane with variance proportional to the cube of the thermal noise. A celestial point source is coherent across the array and moves all bispectra to positive real values proportional to the cube of its brightness \citep{1989AJ.....98.1112K,1995AJ....109.1391R}. This makes the bispectrum useful for the statistical detection of a source anywhere in the field of view.

An interferometer can detect sources with a low signal to noise ratio (SNR) per baseline, $s$, by coherently combining signals over many baselines. For a homogeneous array of $n_a$ elements, there are $n_{bl}=n_a(n_a-1)/2$\ independent baselines, so coherently combining all baselines improves SNR to $s \, \sqrt{n_{bl}}$. For combinations of three antennas, there are $n_{tr}=n_a(n_a-1)(n_a-2)/6$\ closed possibilities. It can be shown that some triple phases are redundant (e.g., $\phi_{123} = \phi_{124} + \phi_{143} + \phi_{234}$), which leaves only $(n_a-1) \, (n_a-2)/2$ ``unique'' triples \citep{1995AJ....109.1391R}. However, the bispectrum is subject to different correlated noise properties, since it is formed by multiplying complex visibilities rather than summing scalar phases. As a result, in the limit of small $s$, all $n_{tr}$\ triples are independent \citep{1989AJ.....98.1112K} and can be coherently combined to improve SNR.

Interferometers can detect a source in many ways. Coherent beamforming sums complex visibilities with weights to direct the beam \footnote{In this context, one can think of imaging as coherently beamforming all possible beams.}. Antenna-based incoherent beamforming sums the power detected at each antenna, while baseline-based beamforming sums power detected on each baseline \citep[i.e., the amplitude of a visibility;][]{tms}. In all of these cases, the SNR of a source detection can be written in terms of the SNR per baseline and size of the array \citep{1989AJ.....98.1112K,1995AJ....109.1391R}. The SNR for coherent beamforming, bispectrum, antenna-based incoherent beamforming, and baseline-based incoherent beamforming are:
\begin{eqnarray}
\rm{SNR}_{bfc}  & = & s \, \sqrt{n_{bl}} , \nonumber \\
\rm{SNR}_{bisp} & = & \frac{1}{2} \, s^3 \, \sqrt{n_{tr}} , \nonumber \\
\rm{SNR}_{bfia} & = & s \, \sqrt{n_a} , \, \nonumber \\
\rm{SNR}_{bfib} & = & \frac{1}{2} \, \frac{s^2}{\sqrt{1 + s^2}} \, \sqrt{n_{bl}}
\label{snreqn}
\end{eqnarray}
\noindent Note that for constant sources, the relation for SNR$_{bisp}$\ is only valid when $s$\ is smaller than 1 \citep{1989AJ.....98.1112K}. However, for transient sources the noise is measured off pulse. This removes the correlated noise effect and makes the $s^3$\ scaling valid for all $s$. We refer to these relations as the ``apparent SNR'', since they are valid when the statistics are measured relative to the standard deviation off source. For Gaussian statistics, the apparent SNR is related to the false positive rate as $\frac{1}{2} * \left(1 - \rm{erf}\left(\frac{SNR}{\sqrt{2}}\right)\right)$ \citep{1995AJ....109.1391R}; a $5\sigma$\ deviation corresponds to a false-positive rate of $2.87\times10^{-7}$.

\begin{figure}[tb]
\centering
\includegraphics[width=0.5\textwidth,bb=10 180 600 600,clip]{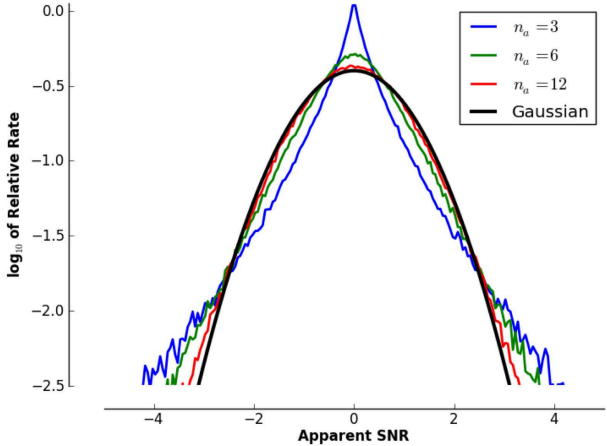}
\caption{Simulated histograms of the noise distribution of the real part of the mean bispectrum. Each bispectrum is calculated by multiplying random complex visibilities for an array of $n_a$\ antennas producing $n_{bl}$\ independent visibilities. The real and imaginary parts of the visibilities are sampled from a Gaussian with standard deviation of 1 and mean of 0. The mean bispectrum for $n_a=3$\ (the smallest array with a bispectrum) is centrally peaked with long tails. For larger arrays, the distribution of the mean bispectrum narrows and approaches that of a Gaussian, as expected from the Central Limit Theorem. The distribution can also be used to estimate the false-positive rate for a signal of a given apparent SNR. 
\label{disn}}
\end{figure}

An important complication to this SNR analysis is that an individual bispectrum measurement is not Gaussian distributed. Visibilities are Gaussian distributed, so multiplying three, noise-like complex visibilities together produces a long-tailed distribution. The product of three independent \emph{real}-valued random variables follows the Meier G function G$^{3,0}_{0,3}$ \citep{lomnicki}; we are not aware of an analytic expression for the real part of the product of three \emph{complex} random variables. Regardless of the exact distribution, the Central Limit Theorem predicts that the mean of many, independent bispectra will approach a Gaussian distribution \citep{1995AJ....109.1391R}. In the presence of pure noise, each bispectrum in an array is independent \citep{1989AJ.....98.1112K}, so the distribution of the mean bispectrum will approach a Gaussian as the number of elements in the array grows, as shown in Figure \ref{disn}.

To include this effect in our sensitivity calculation, we simulated noise-like visibilities and directly measured the flux limit of the coherent beamforming and bispectrum statistics as a function of array size, $n_a$. The flux limit is quantified by calculating $s$\ (from Equations \ref{snreqn}) that would produce the simulated 5$\sigma$\ false-positive rate under Gaussian statistics. This simulation shows that the flux limit of coherent beamforming matches the expectation for a Gaussian-distributed quantity. However, the mean bispectrum has a larger false-positive rate than expected for an apparent 5$\sigma$\ deviation, especially for small $n_a$ where the Central Limit Theorem has less effect. We use these simulations to define an effective bispectrum 5$\sigma$\ flux limit that produces the Gaussian false-positive rate. The effective 5$\sigma$\ flux limit is higher than predicted by Gaussian statistics at a level of 53\%, 26\%, 13\%, and 6\% for $n_a=3, 6, 12, 24$, respectively. For this work, we assume the mean bispectrum is Gaussian distributed, which is reasonable for arrays larger than a roughly a dozen elements for a 5$\sigma$\ flux limit. Future work will develop a more general expression for the false-positive rate and optimizations for transient detection in the non-Gaussian domain.

\begin{figure}[tb]
\centering
\includegraphics[width=0.55\textwidth,bb=30 180 530 550,clip]{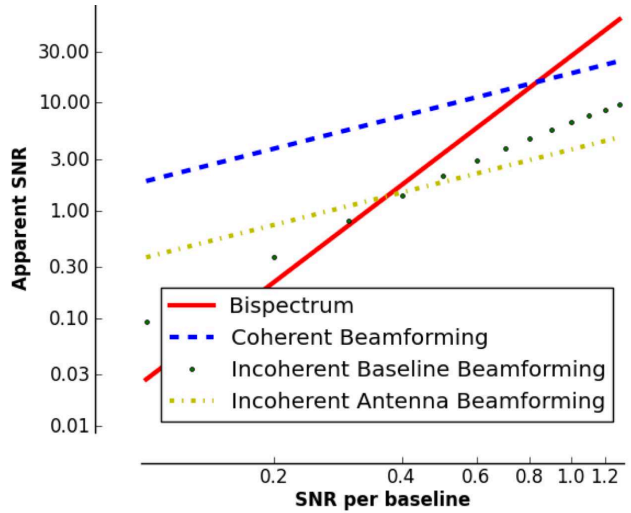}

\includegraphics[width=0.55\textwidth,bb=30 180 530 550,clip]{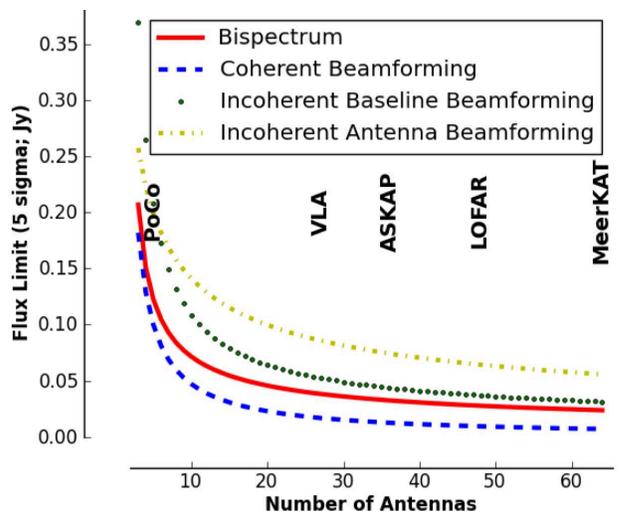}
\caption{\emph{Top:} Apparent SNR for coherent beamforming, bispectrum, and incoherent beamforming techniques as a function of SNR per baseline for a 27-element array. For an effective SNR, defined according to the false-positive rate, the bispectrum SNR never exceeds that of coherent beamforming. \emph{Bottom:} The effective 5$\sigma$\ flux limits as a function of the number of antennas. The effective bispectrum flux limit includes a correction due to non-Gaussianity. The text marks the sizes of major new and planned observatories, including LOFAR \citep{2011AA...530A..80S}, MeerKAT \citep{2009arXiv0910.2935B}, and ASKAP \citep{2008ExA....22..151J}. This assumes Stokes I detection with a SEFD of 400 Jy, 1 GHz bandwidth, and 10 ms integrations, which gives a 1$\sigma$\ sensitivity of 63 mJy per baseline. For reference, the VLA 5$\sigma$\ 10-ms sensitivity is 17 mJy for coherent beamforming, 38 mJy for the bispectrum, 50 mJy for baseline incoherent, and 85 mJy for antenna incoherent detection.
\label{snrtheory}}
\end{figure}

Figure \ref{snrtheory} shows how the apparent SNR of four techniques scales with SNR per baseline for a 27-element array like the VLA. The apparent bispectrum SNR scales as $s^3$, giving it a strong response to bright sources and very little response to faint sources. However, the effective bispectrum SNR, defined according to a Gaussian false-positive rate, would be less sensitive for large $s$. Our simulations suggest that the effective bispectrum SNR never exceeds that of coherent beamforming.

The bottom panel of Figure \ref{snrtheory} shows the effective 5$\sigma$\ flux limit for an array with $n_a$\ elements and sensitivity like the VLA. For the beamforming techniques, this plot essentially shows how the lines in the top panel intersect $\rm{SNR}=5$\ as the array size changes. For the bispectrum, the effective flux limit is shown by correcting for sensitivity loss due to non-Gaussianity, as measured in our simulations. Generally, since large arrays are sensitive to pulses with SNR per baseline much less than one, coherent beamforming and imaging become more sensitive than the bispectrum as $n_a$\ increases. For large $n_a$, the bispectrum and incoherent beamforming techniques scale similarly ($s\propto n_a^{-1/2}$), so the bispectrum is always more sensitive than the incoherent techniques.

To interpret the bispectrum as the brightness of a transient requires subtracting the signature of constant emission from the visibilities. Proper subtraction requires that the visibility fringe does not change over background subtraction time. The maximum fringe rate is equal to the product of the rate of change of the longest baseline in the (\emph{u,v}) plane and the largest direction cosine. In units of radians per day, the maximum fringe rate is $\pi \, b_{\rm{max}} \, \theta_{\rm{f.o.v.}} / \lambda$, where $b_{\rm{max}}$\ is the longest baseline, $\theta_{\rm{f.o.v.}}$\ is the field of view in radians, and $\lambda$\ is the wavelength.
More precisely, the error from subtracting visibilities leaves residual errors proportional to the brightness of sources in the field. \citet{tms} quantify the effect of time smearing as a reduction in the peak brightness of a source. For small changes, we can parametrize this in terms of a 1.4 GHz VLA A configuration observation:
\begin{equation}
\frac{S^p_o - S^p}{S^p_o} = 6 \times 10^{-4} \left[ \left( \frac{\tau}{1 \rm{s}} \right) \, \left( \frac{\theta_{\rm{f.o.v}}}{0\ddeg5} \right) \, \left( \frac{b_{\rm{max}}}{36 \rm{km}} \right) \right]^2
\label{bgerr}
\end{equation}
\noindent where $\tau$\ is the integration time scale.

For visibility (or image) subtraction, the peak flux lost from time smearing is similar to the flux residual in the differenced data given by Equation \ref{bgerr}. This error will contribute to the mean bispectrum in proportion to the fraction of all triples that contain the long baseline. A single antenna is part of $(n_a-1) \, (n_a-2)/2$\ triples, so errors for a long baseline are suppressed by the fraction of triples with that baseline, or a factor of $3/n_a$. Table \ref{sub} gives the longest time scale for which the bispectrum technique can be used on several radio interferometers. The time scale shows when the systematic errors introduced by background subtraction biases the mean bispectrum more than its thermal noise (1$\sigma$). This shows that most interferometers are suitable for using the bispectrum technique to search for transients as long as 1 second. The VLBA is so extended that the fringe rate can only be subtracted on millisecond time scales\footnote{While this estimate uses the full field of view, the VLBA has a much smaller effective field of view due to the delay beam \citep{2011PASP..123..275D}.}.

The bispectrum also contains information about the spatial distribution of emission. For a point source with large $s$, the standard deviation of complex bispectra is $\sigma_B=\sqrt{3} \, s^2 \, Q^3$, where $Q$\ is the noise per baseline \citep{1989AJ.....98.1112K}. Note that the standard deviation of bispectra \emph{is} subject to correlated noise, as opposed to the standard deviation of the mean bispectrum in time. In contrast with a celestial source, interference is often in the near-field or subject to multipath propagation, which produces a random triple phase and a larger variance between bispectra. Another way to think about this is that terrestrial interference can look like a spatially-extended transient. The variance between bispectra provides a simple way to quantify whether a transient is point-like, as expected.

\section{Demonstration}
\label{demo}
To use the bispectrum for transient detection, we assume that the real part of the mean of all bispectra is related to the brightness of a single source in the field of view. To use the bispectrum to detect and localize pulses, we propose the following algorithm:

\begin{itemize}
 \item {\bf Subtract the mean visibility in time:} Differencing visibilities removes all constant emission over some time scale, $\tau$.
 \item {\bf Dedisperse visibilities:} The frequency-dependent arrival time must be removed to maximize SNR per baseline before creating the bispectra.
 \item {\bf Calculate mean and standard deviation of bispectra:} The mean of bispectra gives a single value related to the significance of the transient, while the standard deviation is related to its spatial structure. Both the mean and standard deviation have dimensions of DM and time, so they can be used to set thresholds just as for a single-dish transients search.
 \item {\bf Calibrate and image candidate pulses:} Bispectra can be used to make an image, but calibrated visibilities are needed for an accurate localization \citep{1983AJ.....88..688S}. Visibilities need to be buffered long enough to capture the dispersive delay across the band. The pulse is then imaged in a separate process.
\end{itemize}

\begin{figure}[!tb]
\centering
\includegraphics[width=0.47\textwidth,bb=30 170 550 580,clip]{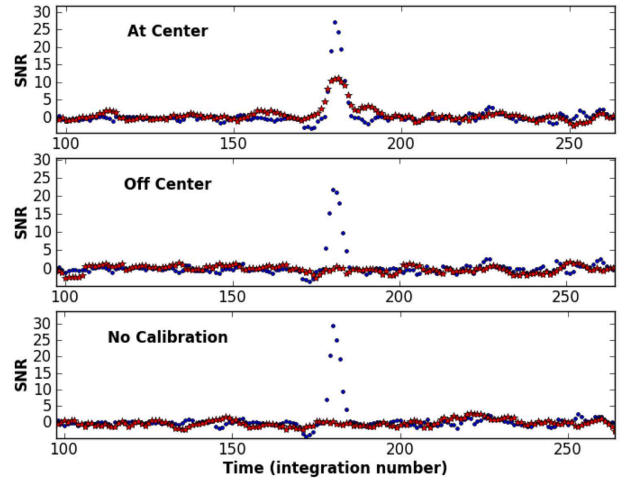}
\caption{\emph{Top:} Time series of the apparent SNR measured by the bispectrum (\emph{blue}) and with coherent beamforming (\emph{red}) in 5-antenna PoCo data of pulsar B0329+54. The PoCo data have a time resolution of 1.2 ms and have been dedispersed assuming the known DM of B0329+54. This panel shows calibrated data with B0329+54 at the phase center. \emph{Middle:} The same plot but with the pulsar off the phase center. \emph{Bottom:} The same plot but with no phase calibration. \label{snr}}
\end{figure}

Figure \ref{snr} shows how the bispectrum and coherent beamforming techniques detect pulses in PoCo data toward B0329+54. For both techniques, the plot shows the apparent pulse SNR when the pulsar is at the phase center, away from the phase center, and at the phase center but uncalibrated. The beamforming technique only detects the pulse when it is at the phase center and data are calibrated. The bispectrum detects the pulse regardless of its location or the phase calibration.

\begin{figure}[tb]
\centering
\includegraphics[width=0.45\textwidth,bb=30 170 550 580,clip]{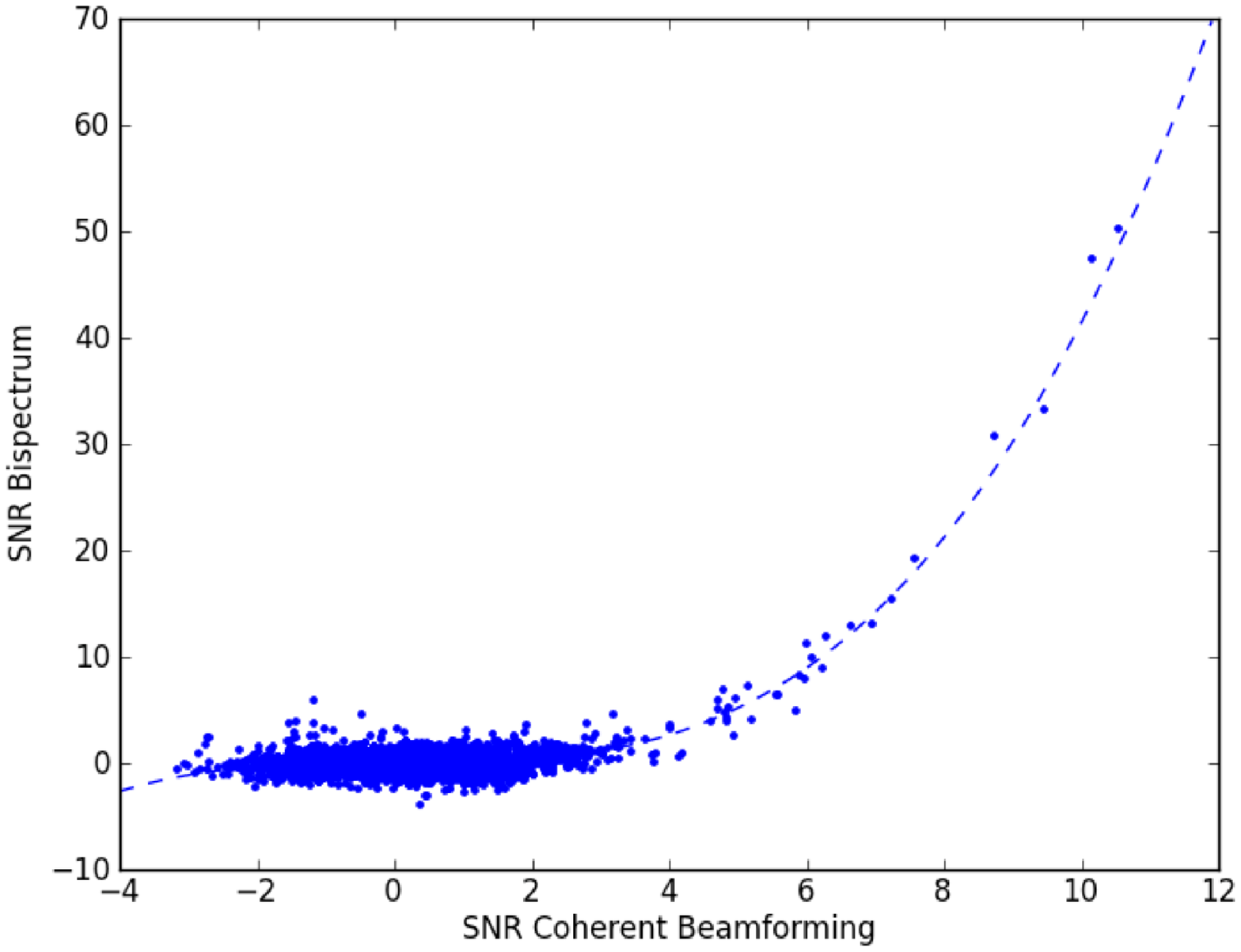}

\includegraphics[width=0.45\textwidth,bb=30 170 550 580,clip]{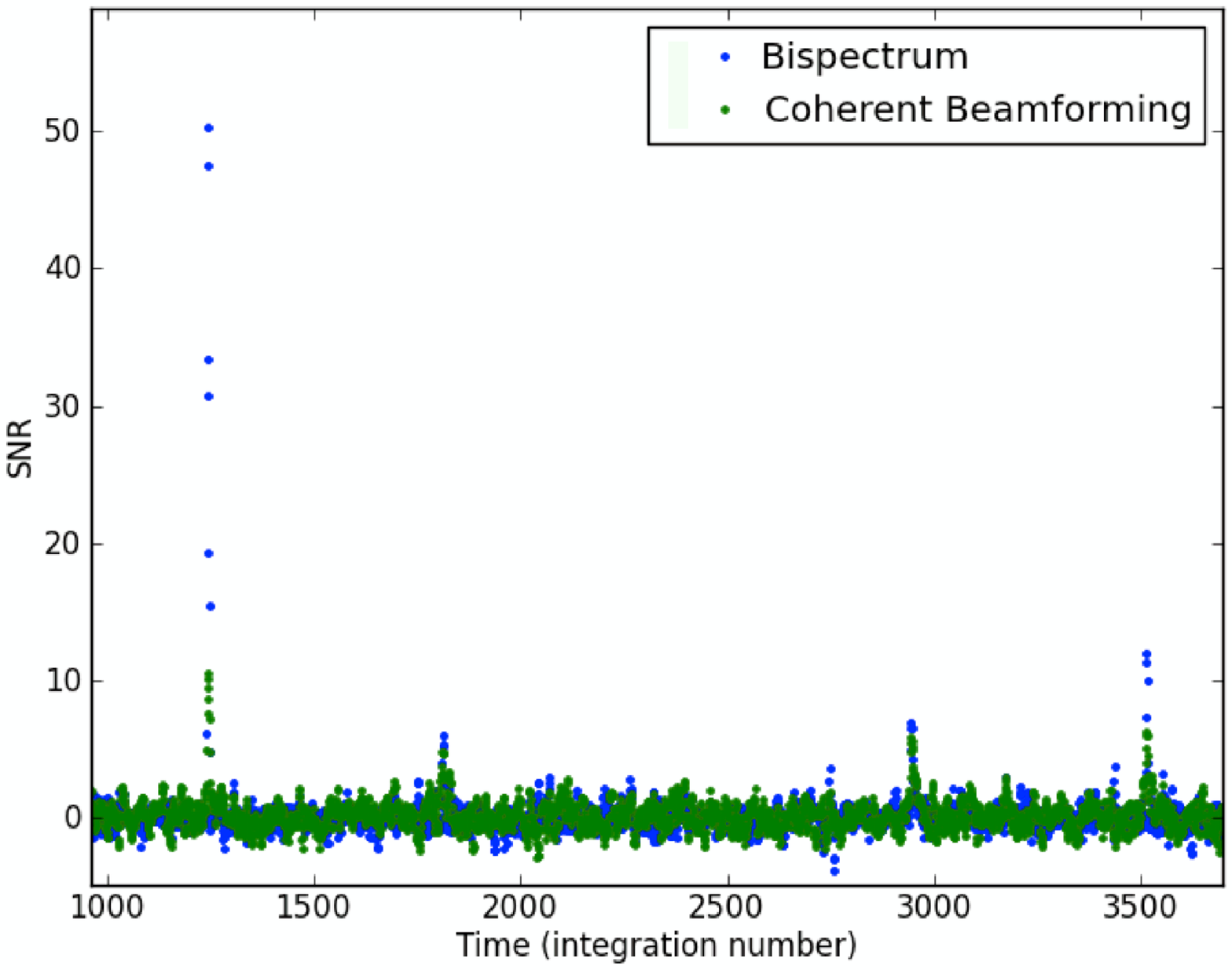}
\caption{\emph{Top:} Apparent SNR of the bispectrum versus that of coherently beamforming of PoCo data toward B0329+54. The SNR is shown for each integration, regardless of significance. The dashed line shows the expected cubic dependence in the apparent SNR, as described in \S \ref{theory} and shown in \citet{1995AJ....109.1391R}. \emph{Bottom:} A time series of the same two statistics for the same data. The rotation period of B0329+54 is about 600 integrations and four of the five pulses in this window are detected. For all detected pulses, the apparent bispectrum SNR is higher than that of beamforming, as expected for a small array as PoCo was used at the ATA. \label{snrobs}}
\end{figure}

The apparent SNR of the bispectrum and coherent beamforming techniques for data including several pulses are shown in Figure \ref{snrobs}. The top panel in Figure \ref{snrobs} shows that the PoCo pulse SNR distribution measured with beamforming and the bispectrum follow the relation of \citet{1995AJ....109.1391R} well. Since there are only nine baselines and five triples in this analysis, the pulses shown here have relatively high $s$. The brightest pulse in this data has $s\sim3$, corresponding to a brightness of about 50 Jy. A similar search with a larger array would have a much slower rise in $\rm{SNR}_{bisp}$\ relative to $\rm{SNR}_{bfc}$.

\begin{figure}[tb]
\centering
\includegraphics[width=0.48\textwidth,bb=0 170 530 580,clip]{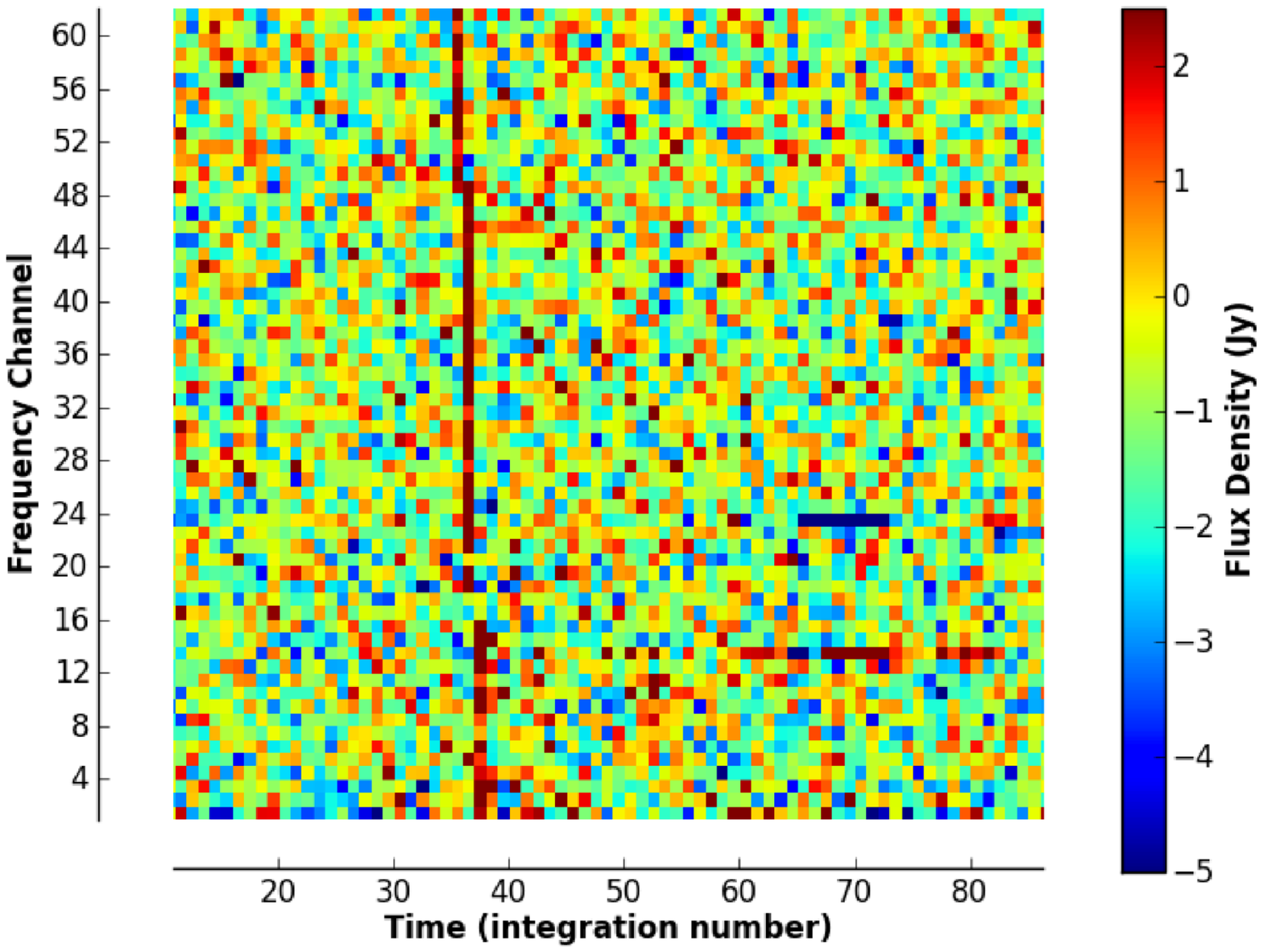}

\includegraphics[width=0.48\textwidth,bb=0 170 560 580,clip]{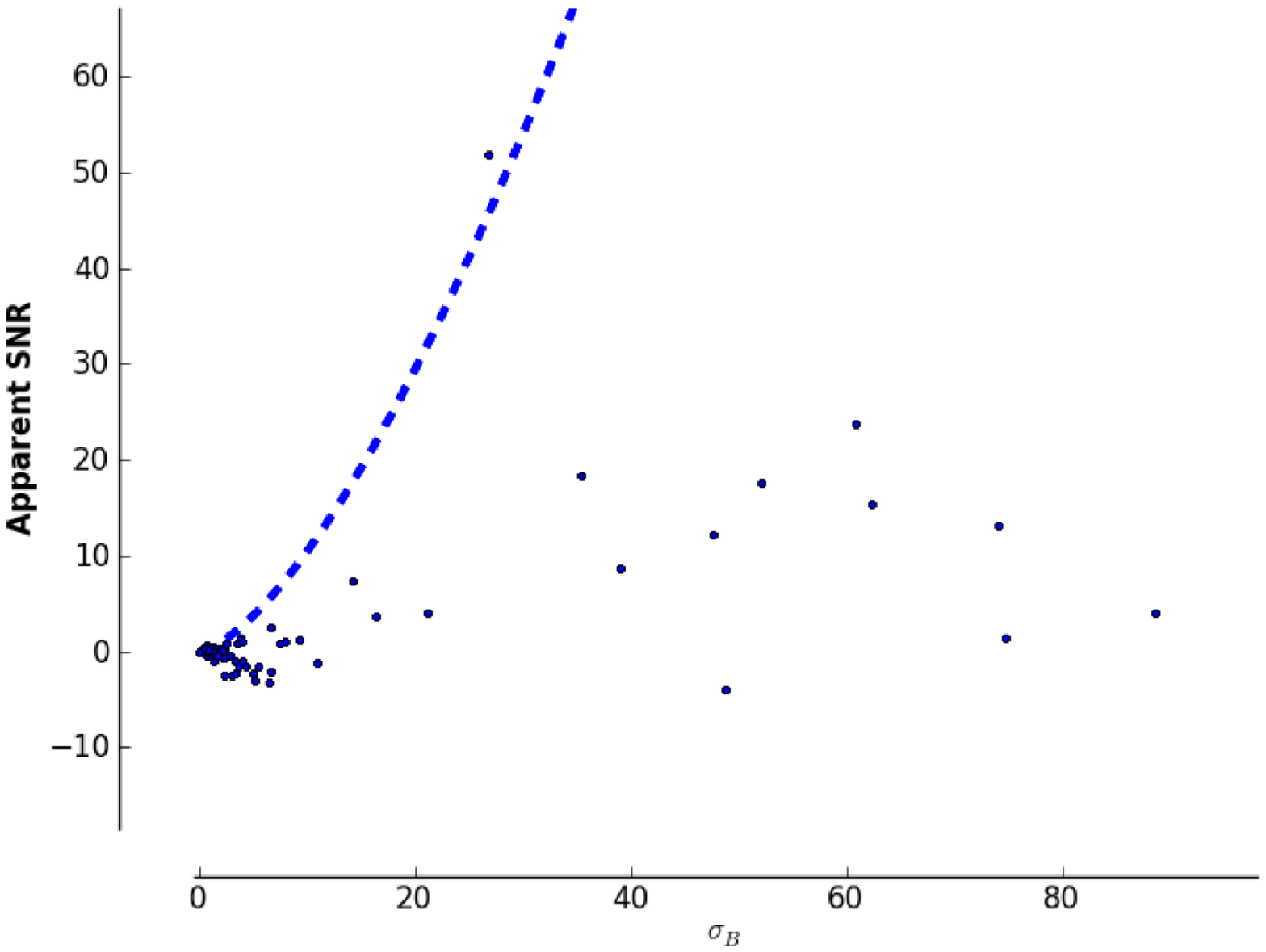}
\caption{\emph{Top:} VLA dynamic spectrum toward the Crab pulsar at 1.4 GHz with 12 ms integrations and 128 MHz bandwidth. The dynamic spectrum is formed from the sum of all cross correlations, effectively forming a synthesized beam at the location of the Crab pulsar with a brightness in units of Janskys. At the left is a pulse from the Crab pulsar, which appears as a red vertical line with a slight kink due to dispersion. Two channels show interference of similar strength toward the right. \emph{Bottom:} The apparent bispectrum SNR versus standard deviation of all bispectra for each dedispersed integration. The blue dashed line shows the theoretical expectation for a point source with $s\gg1$. The one point near that line corresponds to the Crab pulse with $s\approx4$. \label{rfi}}
\end{figure}

Figure \ref{rfi} shows how the bispectrum can distinguish between a celestial point transient and terrestrial interference. The top panel shows a typical dynamic spectrum for a VLA observation of the Crab pulsar with 12 millisecond integrations. As is well known from single-dish observing, interference is bright enough to be detected even after dedispersion. The bottom panel shows the mean bispectrum and standard deviation over bispectra for the same pulse and interference. The interference has a large standard deviation of bispectra, while the true Crab pulse follows the expected theoretical relation for $s\gg1$.

\section{Data and Operations Rates}
\label{demand}
Another strength of the bispectrum technique is its computationally efficiency. Table \ref{comp} shows how the computational demand scales for three coherent fast transient detection techniques at the VLA. These computational demand values are scaled to operations per second (flops) by multiplying by the integration rate in Hz. For the bispectrum and imaging techniques, the visibilities are dedispersed for each baseline and DM trial, while coherent beamforming is traditionally done before dedispersion. The imaging FFT requires $N_{\rm{bl}} \, N_{\rm{ch}} + 5 \, N_{\rm{pix}}^2 \, \rm{log}_{2} N_{\rm{pix}}^2$\ operations per image \citep{1999ASPC..180..127B}. For the VLA D configuration, images require roughly 128 pixels per side, so $N_{\rm{pix}}=128$.

The bispectrum technique is the most computationally efficient technique for coherent pulse detection. Running this algorithm on 10 ms integrations at the VLA would require roughly 34 Gflops. Imaging is at least two times more demanding (this estimate excludes image cleaning). The coherent beamforming is driven by the number of beams needed to fill the field of view. Even in the compact D configuration, roughly 7000 beams are needed. The VLA A configuration would require about 900 times more synthesized beams and 3 Pflops to calculate. The large computational demand of coherent beamforming explains why it is typically only done with a few beams \citep{2011AA...530A..80S}.

Our pulse detection algorithm consists of many independent processes (across frequency, DM trial, and time), so the problem is readily parallelized. Moreover, aspects of the algorithm, such as the large number of multiplications and dedispersion, are well suited to GPUs \citep{2011MNRAS.417.2642M,2011arXiv1107.4264C}. Table \ref{comp} suggests that the bispectrum computational demand can be met with a single GPU-accelerated server. However, the data rate associated with millisecond integrations from a large array can limit its application for some correlator designs, so careful attention must be paid to data rate bottlenecks. For example, at the VLA, the visibility integration time must be greater than roughly 10 ms, since nodes of the WIDAR compute cluster have 1 GbE input ports \citep{2011ApJ...739L...1P}. A correlator that can produce visibilities at high ($\ge 10$\ Gb s$^{-1}$) data rates is more likely to be limited by computational resources, driven largely by the need to dedisperse visibilities from each baseline. In this case, it will be easier to run this algorithm in real time on arrays with a small number of elements.

This description has focused on how the bispectrum can detect an astrophysical transient, but it can't be used to localize it. While the bispectrum can be used to make images of the sky \citep{1983AJ.....88..688S}, precise localization requires external phase calibration. Thus, the final step in this algorithm is to save candidate transient data for calibration and imaging using traditional techniques \citep{1999ASPC..180..127B}. This kind of post-processing is computationally more intensive than the bispectrum and is not feasible in real time on every integration. However, since we only need to calibrate and image the occasional candidate event, the total computational demand will still be dominated by the steps before the event detection.

\section{Conclusion}
We demonstrate how to use the bispectrum to detect and localize transients with radio interferometers. The technique produces a single statistic proportional to the brightness of celestial transients, thus reducing the problem of interferometric pulse detection to that of single-dish pulse detection. For a small cost in sensitivity, this technique makes an interferometer into a massive multibeam receiver with a proportionately large survey speed. Using coherent beamforming to survey a similar area with next-generation radio interferometers would require $10^2$\ to $10^5$\ times more computational power.

The bispectrum technique is efficient enough to build into a correlator for real-time, millisecond transient detection. This system could save visibility data associated with a candidate and apply standard calibration and imaging techniques to localize it to arcsecond precision. The growing capacity of correlators for interferometers will make it possible to run such a system in real time to commensally search all data for transients. The technique can also be used to find transients offline and on longer time scales. In that case, the bispectrum is useful because it is computationally simple, resistant to interference, and requires no calibration.

A real-time, commensal system would probe an unprecedented volume of data, giving it sensitivity to rare transients. Each pulse could be localized in real time to arcsecond precision, opening a range of new science. Pulses from rotating radio transients \citep{2006Natur.439..817M} could be associated with X-ray counterparts to help address their relation to the normal pulsar population. The dispersion of extragalactic transients would measure the electron density of the intergalactic medium for the first time \citep{2003ApJ...596..982M,2007Sci...318..777L}; localization will help find optical counterparts and redshifts to host galaxies for each pulse. A fast transients survey of the Galactic center could detect individual pulses from pulsars that would probe the dispersion toward the region and potentially find a pulsar in orbit around a supermassive black hole \citep{2004ApJ...615..253P,2010ApJ...715..939M,2011arXiv1111.4216W}. Magnetospheres of exoplanets and active stars can emit bright, complex bursts that could uniquely identify them \citep{2004ApJ...612..511L,2006ApJ...653..690H}. 

The strength of the bispectrum technique is its real-time ability to find transients throughout the field of view and reduce the flow of data to select times. This shows how algorithms will be key to efficiently extracting the best science in the era of big data. The ability to find pulses with interferometers may have application to communications problems, such as developing smart antennas for the efficient use of spectrum. If so, this technique would follow in the footsteps of previous transients searches that contributed to the development of Wi-Fi \citep{1978Natur.276..590O}.

\acknowledgements{We thank Michael Rupen, Rick Perley, and Bryan Butler for providing VLA fast dump data, Peter Williams for developing Miriad-Python, and Gregg Hallinan, Scott Vander Wiel, and Olaf Wucknitz for useful discussion. We are grateful to National Science Foundation grant AST-0807151 for support of this research. This research has made use of NASA's Astrophysics Data System Bibliographic Services.}

{\it Facility:} \facility{ATA, VLA}


\clearpage

\begin{deluxetable}{cccc|c}
\tablecaption{Time Scale for Using Bispectrum Technique on Radio Interferometers \label{sub}}
\tablewidth{0pt}
\tablehead{
\colhead{Telescope} & \colhead{$\theta_{\rm{f.o.v.}}$} & \colhead{$b_{\rm{max}}$} & \colhead{Error}\footnote{Largest error expected in mean bispectrum, assuming Cassiopeia A located at the edge of the field of view \citep{1977AA....61...99B}.} & \colhead{$\tau_{\rm{max}}$} \\
                    &  (deg)                         &   (km)                  & (mJy)                                   & (s) 
}
\startdata
VLA-A\footnote{Measured VLA sensitivity in A configuration at 1.4 GHz with 1 GHz of bandwidth.}   & 0.5 & 36  &   5  & 0.2 \\
VLA-D\footnote{Measured VLA sensitivity in D configuration at 1.4 GHz with 1 GHz of bandwidth.}   & 0.5 & 1  &   5  & 7.2 \\
VLBA\footnote{Measured VLBA sensitivity using a 512 MHz bandwidth.}   & 0.5 & 8611&  70  & 0.003  \\
PAPER\footnote{Measured PAPER sensitivity at 150 MHz of 100 mJy in 5 seconds with coherent beamforming of new 64-element array \citep{2010AJ....139.1468P}.}   & 40  & 0.3 & 215  & 1.6 \\
LOFAR\footnote{Predicted LOFAR sensitivity at 150 MHz of 10 mJy in 1 second with 20-station coherent sum of core stations \citep{2011AA...530A..80S}.}   & 2.4 &  1  &  40  & 2 \\
MeerKAT\footnote{Predicted MeerKAT ``Phase 2'' sensitivity with 80 dishes, 2 GHz bandwidth, and 30 K system temperature \citep{2009arXiv0910.2935B}.} & 1   & 20  &   7  & 0.2 \\
ASKAP\footnote{Predicted ASKAP ``strawman'' sensitivity with 30 elements, 300 MHz bandwidth, and 50 K system temperature \citep{2008ExA....22..151J}.}   & 3   &  2  &  18  & 1 \\
\enddata
\end{deluxetable}

\begin{deluxetable}{c|cc|cc|cc}
\tablecaption{Computational Demand per Integration for VLA Fast Transient Search Algorithms\tablenotemark{a} \label{comp}}
\tablewidth{0pt}
\tabletypesize{\scriptsize}
\tablehead{
\colhead{Process} & \multicolumn{2}{|c|}{Bispectrum} & \multicolumn{2}{c|}{Coherent Beamforming} & \multicolumn{2}{c}{Imaging} \\ 
        &      Scaling &   Demand   & Scaling  & Demand  & Scaling  & Demand
}
\startdata
Subtraction            & $10 N_{\rm{bl}} N_{\rm{pol}} N_{\rm{ch}} N_{\rm{t}}$ & 14 Mop & $10 N_{\rm{bl}} N_{\rm{pol}} N_{\rm{ch}} N_{\rm{t}}$ & 14 Mop & $10 N_{\rm{bl}} N_{\rm{pol}} N_{\rm{ch}} N_{\rm{t}}$ & 14 Mop \\ 
Beamform               & -- & -- & $3 N_{\rm{b}} N_{\rm{bl}} N_{\rm{pol}} N_{\rm{ch}} N_{\rm{t}}$ & 28 Gop & -- & -- \\
Dedisperse             & $N_{\rm{bl}} N_{\rm{pol}} N_{\rm{ch}} N_{\rm{DM}} N_{\rm{t}}$ & 287 Mop & $N_{\rm{b}} N_{\rm{ch}} N_{\rm{DM}} N_{\rm{t}}$ & 2.8 Gop & $N_{\rm{bl}} N_{\rm{pol}} N_{\rm{ch}} N_{\rm{DM}} N_{\rm{t}}$ & 287 Mop \\
Image\footnote{Excluding image cleaning, source identification, and application of calibration.}           & -- & -- & -- & -- & $1.5e6 N_{\rm{DM}} N_{\rm{t}}$ & 602 Mop \\
Bispectrum             & 16 $N_{\rm{tr}} N_{\rm{pol}} N_{\rm{DM}} N_{\rm{t}}$ & 37 Mop & -- & -- & -- & -- \\ \hline 
Total\footnote{Assuming high-demand case: $N_{\rm{bl}}=351$, $N_{\rm{tr}}=2925$, $N_{\rm{SB}}=16$, $N_{\rm{pol}}=2$, $N_{\rm{ch}}=1024$, $N_{\rm{b}}=7000$, and $N_{\rm{DM}}=200$. This produces spectra with 2 GHz of bandwidth and 2 MHz channel size. The number of dispersion trials, $N_{\rm{DM}}=200$, is appropriate for a high-demand case of $DM=1000$\ pc cm$^{-3}$ for 1.2--2.0 GHz band and 10 ms integration time. The number of time scales probed is effectively $N_{\rm{t}}=2$, assuming that the number of trials scales as the inverse of the time scale. Imaging and beamforming assume the D configuration, which has a longest baseline of 1 km. The A configuration requires 900 times more pixels and beams.}                  & & 339 Mop & & 33 Gop & & 904 Mop
\enddata

\end{deluxetable}

\end{document}